\documentclass[conference]{IEEEtran} 
\IEEEoverridecommandlockouts
\usepackage{cite}
\usepackage{amsmath,amssymb,amsfonts}
\usepackage{algorithmic}
\usepackage{graphicx}
\usepackage{textcomp}
\usepackage{xcolor}
\usepackage{url}
\def\BibTeX{{\rm B\kern-.05em{\sc i\kern-.025em b}\kern-.08em
    T\kern-.1667em\lower.7ex\hbox{E}\kern-.125emX}}
\begin{document}

\title{Anonymized Network Sensing Graph Challenge
\thanks{
Research was sponsored by the Department of the Air Force Artificial Intelligence Accelerator and was accomplished under Cooperative Agreement Number FA8750-19-2-1000. The views and conclusions contained in this document are those of the authors and should not be interpreted as representing the official policies, either expressed or implied, of the Department of the Air Force or the U.S. Government. The U.S. Government is authorized to reproduce and distribute reprints for Government purposes notwithstanding any copyright notation herein.
}
}

\author{\IEEEauthorblockN{
Hayden Jananthan$^1$, Michael Jones$^1$,  William Arcand$^1$, David Bestor$^1$, William Bergeron$^1$, Daniel Burrill$^1$, \\ Aydin Buluc$^2$, Chansup Byun$^1$, Timothy Davis$^3$, Vijay Gadepally$^1$, Daniel Grant$^4$, Michael Houle$^1$, Matthew Hubbell$^1$, \\ Piotr Luszczek$^{1,5}$, Peter Michaleas$^1$, Lauren Milechin$^1$, Chasen Milner$^1$, Guillermo Morales$^1$, Andrew Morris$^4$, \\ Julie Mullen$^1$, Ritesh Patel$^1$, Alex Pentland$^1$, Sandeep Pisharody$^1$, Andrew Prout$^1$,  Albert Reuther$^1$, Antonio Rosa$^1$, \\ Gabriel Wachman$^1$, Charles Yee$^1$, Jeremy Kepner$^1$
\\
\IEEEauthorblockA{
$^1$MIT,  $^2$LBNL, $^3$Texas A\&M, $^4$GreyNoise, $^5$University of Tennessee
}}}
\maketitle

\begin{abstract}
The MIT/IEEE/Amazon GraphChallenge encourages community approaches to developing new solutions for analyzing graphs and sparse data derived from social media, sensor feeds, and scientific data to discover relationships between events as they unfold in the field.  The anonymized network sensing Graph Challenge seeks to enable large, open, community-based approaches to protecting networks. Many large-scale networking problems can only be solved with community access to very broad data sets with the highest regard for privacy and strong community buy-in. Such approaches often require community-based data sharing.  In the broader networking community (commercial, federal, and academia) anonymized source-to-destination traffic matrices with standard data sharing agreements have emerged as a data product that can meet many of these requirements.  This challenge provides an opportunity to highlight novel approaches for optimizing the construction and analysis of anonymized traffic matrices using over 100 billion network packets derived from the largest Internet telescope in the world (CAIDA).  This challenge specifies the anonymization, construction, and analysis of these traffic matrices.  A GraphBLAS reference implementation is provided, but the use of GraphBLAS is not required in this Graph Challenge. As with prior Graph Challenges the goal is to provide a well-defined context for demonstrating innovation. Graph Challenge participants are free to select (with accompanying explanation) the Graph Challenge elements  that are appropriate for highlighting their innovations.
\end{abstract}


\section{Introduction}

The MIT/IEEE/Amazon GraphChallenge encourages community approaches to developing new solutions for analyzing graphs and sparse data.  GraphChallenge.org provides a well-defined community venue for stimulating research and highlighting innovations in graph and sparse data analysis software, hardware, algorithms, and systems.  The target audience for these challenges are any individual or team that seeks to highlight their contributions to graph and sparse data analysis software, hardware, algorithms, and/or systems.   The Sparse DNN  \cite{kepner2019sparse, bisson2019gpu, davis2019write, 
lin2020novel, hidayetoglu2020atscale, xin2021fast, sun2022accelerating, xu2022towards, dun2023adaptive},  Stochastic Block Partitioning \cite{kao2017streaming, halappanavar2017scalable,  priest2020scaling, uppal2021faster, wanye2023integrated}, Subgraph Isomorphism \cite{samsi2017static, wolf2017fast, pearce2017triangle, voegele2017parallel, bisson2017static, hu2018high, bisson2018update, yasar2018fast, pearce2018ktruss, pandey2019hindex, blanco2019exploration, pearce2019quadrillion, samsi2020graphchallenge, ghosh2020tric, wang2023smog}, and PageRank \cite{dreher2016pagerank, zhou2017design, sadi2018pagerank} Graph Challenges have enabled a new generation of graph analysis by highlighting the benefits of novel innovations.  Graph Challenge is part of the long tradition of  challenges that have played a key role in advancing computation, AI, and other fields, such as, YOHO~\cite{yoho}, MNIST~\cite{mnist}, HPC Challenge~\cite{dongarra2017hpc}, ImageNet~\cite{imagenet} and VAST~\cite{vast1,vast2}.  More recently, major community activities, such as the NeurIPS conference and the MIT AI Accelerator \cite{gadepally2022developing}, have prioritized the regular development of datasets, benchmarks, and challenges.

\begin{figure}
\center{\includegraphics[width=1.0\columnwidth]{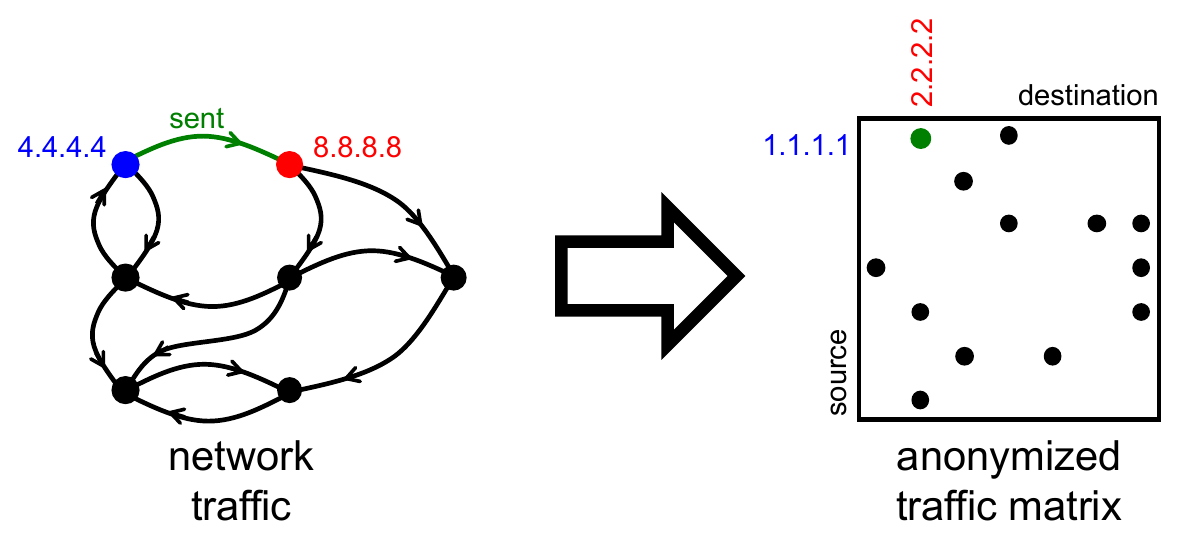}}
      	\caption{{\bf Network Traffic Messages to Anonymized Traffic Matrix.}   Network traffic uses numbers to denote the source and destination addresses of messages.   Network traffic messages can be aggregated and summarized into traffic matrices for analysis.  These traffic matrices, when coupled with data sharing agreements, can be anonymized by relabeling source addresses (e.g., {\sf\textcolor{blue}{4.4.4.4}} $\rightarrow$ {\sf\textcolor{blue}{1.1.1.1}}) and destination addresses (e.g., {\sf\textcolor{red}{8.8.8.8}} $\rightarrow$ {\sf\textcolor{red}{2.2.2.2}})  using various anonymization schemes.  The Anonymized Network Traffic Graph Challenge provides an opportunity to highlight novel approaches for optimizing the construction and analysis of anonymized traffic matrices from network traffic.}
      	\label{fig:NetworkToMatrix}
\end{figure}

The Anonymized Network Sensing Graph Challenge seeks to enable large, open, community-based approaches to protecting networks \cite{atkins2021improvised, atkins2021cooperation, demchak2021achieving, weed2022beyond, weed2023beyond}.   Many large-scale networking problems can only be solved with community access to very broad data sets with the highest regard for privacy and strong community buy-in \cite{kepner2021zero, pisharody2021realizing, pentland2022building}. In the broader networking community (commercial, federal, and academia) anonymized source-to-destination traffic matrices with standard data sharing agreements have emerged as a data product that can meet many of these requirements (see Figure~\ref{fig:NetworkToMatrix}).  This challenge provides an opportunity to highlight novel approaches for optimizing the construction and analysis of anonymized traffic matrices.

A GraphBLAS reference implementation is provided, but the use of GraphBLAS is not required in this Graph Challenge. GraphBLAS anonymized hypersparse traffic matrices represent one set of design choices for analyzing network traffic  \cite{kepner16mathematical, buluc17design,yang2018implementing, kepner2018mathematics, davis2019algorithm, mattson2019lagraph, cailliau2019redisgraph, davis2019write, aznaveh2020parallel, brock2021introduction, pelletier2021graphblas, jones2022graphblas, trigg2022hypersparse, davis2023algorithm}.  Specifically, the use cases requiring some data on all packets (no down-sampling), high performance, high compression,  matrix-based analysis, anonymization, and open standards.  There are a wide range of alternative graph/network analysis technologies and many good implementations  achieve performance close to the limits of the underlying computing hardware  \cite{tumeo2010efficient, kumar2018ibm, ezick2019combining, gera2020traversing, azad2020evaluation, du2021interactive, acer2021exagraph, blanco2021delayed, ahmed2021online, azad2021combinatorial, koutra2021power}.  Likewise, there are many network analysis tools that focus on providing a rich interface to the full diversity of data found in network traffic \cite{hofstede2014flow, sommer2003bro, lucente2008pmacct}.  Each of these technologies has appropriate use cases in the broad field of Internet traffic analysis.

The outline of the rest of the paper is as follows.  First, some basic network quantities  are defined in terms of traffic matrices.   Second, the steps of the Anonymized Network Sensing Graph Challenge and computational metrics are described.  Next, the test data sets both real and random are presented.     Finally, some preliminary performance results of the reference implementation are provided.

\section{Anonymized Network Traffic Matrices}

Network data must be handled with care.  The Center for Applied Internet Data Analysis (CAIDA) based at the University of California's San Diego Supercomputer Center has pioneered trusted data sharing best practices that combine anonymizing source and destination internet addresses using CryptoPAN \cite{fan2004prefix} with data sharing agreements.  These data sharing best practices include the following principles \cite{kepner2021zero}.
\begin{itemize}
\item Data is made available in curated repositories.
\item Using standard anonymization methods where needed: hashing, sampling, and/or simulation.
\item Registration with a repository and demonstration of legitimate research need.
\item Recipients legally agree to neither repost a corpus nor deanonymize data.
\item Recipients can publish analysis and data examples necessary to review research.
\item Recipients agree to cite the repository and provide publications back to the repository.
\item Repositories can curate enriched products developed by researchers.
\end{itemize}

\begin{figure}
\center{\includegraphics[width=1.0\columnwidth]{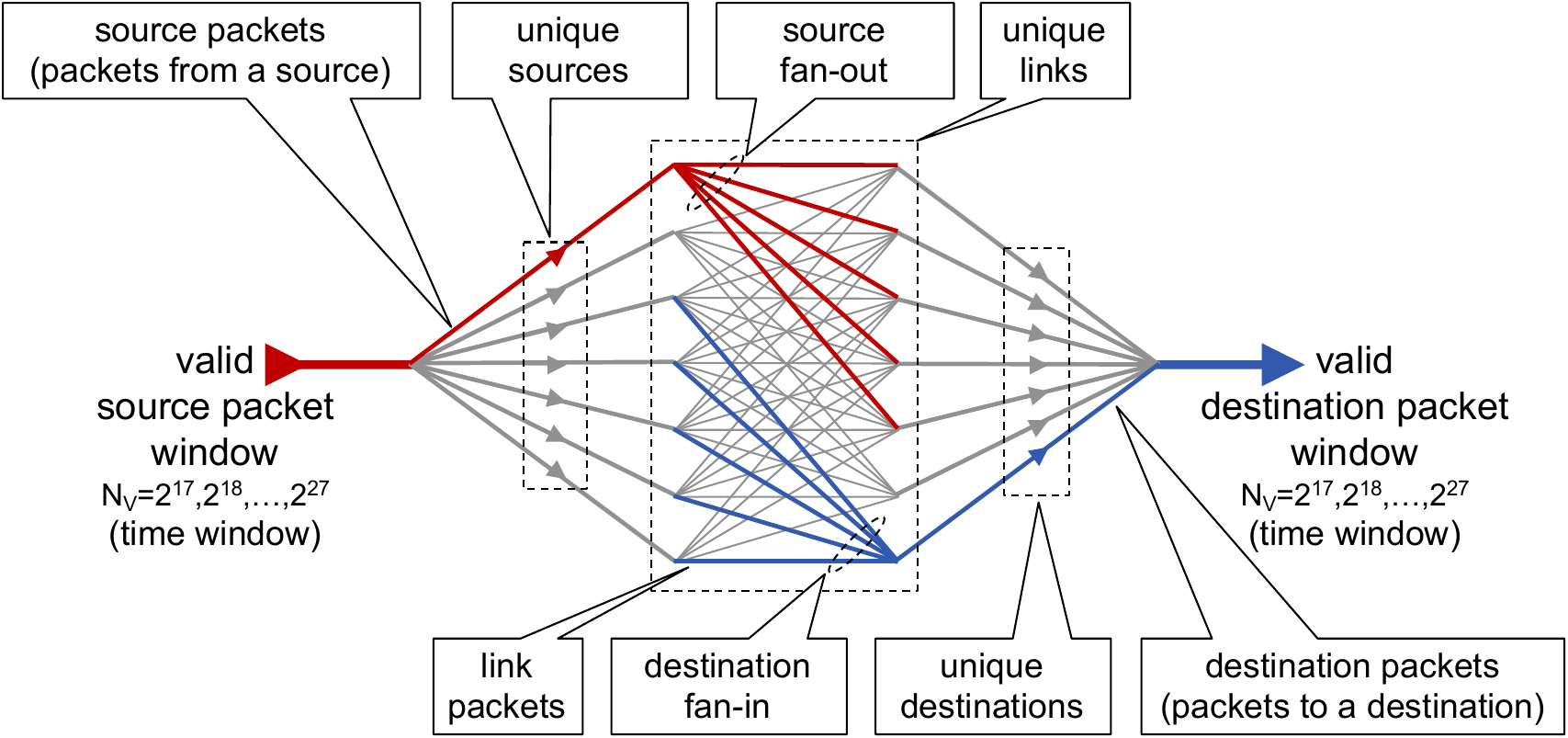}}
      	\caption{{\bf Streaming Network Traffic Quantities.} Internet traffic streams of $N_V$ valid packets are divided into a variety of quantities for analysis: source packets, source fan-out, unique source-destination pair packets (or links), destination fan-in, and destination packets.  Figure adapted from \cite{kepner19hypersparse}.}
      	\label{fig:NetworkDistribution}
\end{figure}
\begin{table}
\caption{Network Quantities from Traffic Matrices}
\vspace{-0.25cm}
Formulas for computing network quantities from a traffic matrix ${\bf A}_t$ at time $t$ in both summation and matrix notation. ${\bf 1}$ is a column vector of all 1's, $^{\sf T}$  is the transpose operation, and $|~|_0$ is the zero-norm that sets each nonzero value of its argument to 1\cite{karvanen2003measuring}.  These formulas are unaffected by matrix permutations and work on anonymized data.  \underline{\smash{Underlined}} quantities are those specified in the anonymized network sensing Graph Challenge. Table adapted from \cite{kepner2020multi}.
\begin{center}
\begin{tabular}{p{1.5in}p{0.9in}p{0.6in}}
\hline
{\bf Aggregate} & {\bf ~~~~Summation} & {\bf ~Matrix} \\
{\bf Property} & {\bf ~~~~~~Notation} & {\bf Notation} \\
\hline
\underline{\smash{Valid packets}} $N_V$ & $~\sum_i ~ \sum_j ~ {\bf A}_t(i,j)$ & $~{\bf 1}^{\sf T} {\bf A}_t {\bf 1}$ \\
\underline{\smash{Unique links}} & $~~\sum_i ~ \sum_j |{\bf A}_t(i,j)|_0$  & ${\bf 1}^{\sf T}|{\bf A}_t|_0 {\bf 1}$ \\
Link packets from $i$ to $j$ & $~~~~~~~~~~~~~~{\bf A}_t(i,j)$ & ~~~$~{\bf A}_t$ \\
\underline{\smash{Max link packets}} ($d_{\rm max}$) & $~~~~~\max_{ij}{\bf A}_t(i,j)$ & $\max({\bf A}_t)$ \\
\hline
\underline{\smash{Unique sources}} & $~\sum_i |\sum_j ~~ {\bf A}_t(i,j)|_0$  & ${\bf 1}^{\sf T}|{\bf A}_t {\bf 1}|_0$ \\
Packets from source $i$ & $~~~~~~~\sum_j ~ {\bf A}_t(i,j)$ & ~~$~~{\bf A}_t  {\bf 1}$ \\
\underline{\smash{Max source packets}} ($d_{\rm max}$)  & $ \max_i \sum_j ~ {\bf A}_t(i,j)$ & $\max({\bf A}_t {\bf 1})$ \\
Source fan-out from $i$ & $~~~~~~~~~~~\sum_j |{\bf A}_t(i,j)|_0$  & ~~~$|{\bf A}_t|_0 {\bf 1}$ \\
\underline{\smash{Max source fan-out}} ($d_{\rm max}$) & $ \max_i \sum_j |{\bf A}_t(i,j)|_0$  & $\max(|{\bf A}_t|_0 {\bf 1})$ \\
\hline
\underline{\smash{Unique destinations}} & $~\sum_j |\sum_i ~ {\bf A}_t(i,j)|_0$ & $|{\bf 1}^{\sf T} {\bf A}_t|_0 {\bf 1}$ \\
Destination packets to $j$ & $~~~~~~~\sum_i ~ {\bf A}_t(i,j)$ & ${\bf 1}^{\sf T}|{\bf A}_t|_0$ \\
\underline{\smash{Max destination packets}} ($d_{\rm max}$) & $ \max_j \sum_i ~ {\bf A}_t(i,j)$ & $\max({\bf 1}^{\sf T}|{\bf A}_t|_0)$ \\
Destination fan-in to $j$ & $~~~~~~~~~~~\sum_i |{\bf A}_t(i,j)|_0$ & ${\bf 1}^{\sf T}~{\bf A}_t$ \\
\underline{\smash{Max destination fan-in}} ($d_{\rm max}$) & $ \max_j \sum_i |{\bf A}_t(i,j)|_0$ & $\max({\bf 1}^{\sf T}~{\bf A}_t)$ \\
\hline
\end{tabular}
\end{center}
\label{tab:Aggregates}
\end{table}

Network traffic data can be viewed as a traffic matrix where each row is a source and each column is a destination (see Figure~\ref{fig:NetworkToMatrix}). A primary benefit of constructing anonymized  traffic matrices is the efficient computation of a wide range of network quantities via matrix mathematics. Figure~\ref{fig:NetworkDistribution} illustrates essential quantities found in all streaming dynamic networks. These quantities are all computable from anonymized traffic matrices created from the source and destination addresses found in Internet packet headers
\cite{soule2004identify, zhang2005estimating, mucha2010community, tune2013internet}.  It is common to filter the Internet Protocol (IP) packets down to a valid set for any particular analysis.   Such filters may limit particular sources, destinations, protocols, and time windows. To reduce statistical fluctuations, the streaming data should be partitioned so that for any chosen time window all data sets have the same number of valid packets \cite{kepner2020multi}.  At a given time $t$, $N_V$ consecutive valid packets are aggregated from the network traffic into a  matrix ${\bf A}_t$, where ${\bf A}_t(i,j)$ is the number of valid packets between the source $i$ and destination $j$. The sum of all the entries in ${\bf A}_t$ is equal to $N_V$:
\begin{equation*}
    \sum_{i,j} {\bf A}_t(i,j) = N_V
\end{equation*}
Constant packet, variable time samples simplify the statistical analysis of the heavy-tail distributions commonly found in network traffic quantities \cite{kepner19hypersparse, nair2020fundamentals, kepner2022new}.  All the network quantities depicted in Figure~\ref{fig:NetworkDistribution} can be readily computed from ${\bf A}_t$ using the formulas listed in Table~\ref{tab:Aggregates}.  Because matrix operations are generally invariant to permutation (reordering of the rows and columns), these quantities can readily be computed from anonymized data.  Furthermore, the anonymized data can be analyzed by source and destination subranges (subsets when anonymized)  using simple matrix multiplication.  For a given subrange represented by an anonymized  diagonal matrix ${\bf A}_r$, where ${\bf A}_r(i,i) = 1$ implies  source/destination $i$ is in the range, the traffic within the subrange can be computed via: ${\bf A}_r {\bf A}_t  {\bf A}_r$. Likewise, for additional privacy guarantees that can be implemented at collection, the same method can be used to exclude a range of data from the traffic matrix:
\begin{equation*}
     {\bf A}_t - {\bf A}_r {\bf A}_t  {\bf A}_r
\end{equation*}

\begin{figure}
\center{\includegraphics[width=1.0\columnwidth]{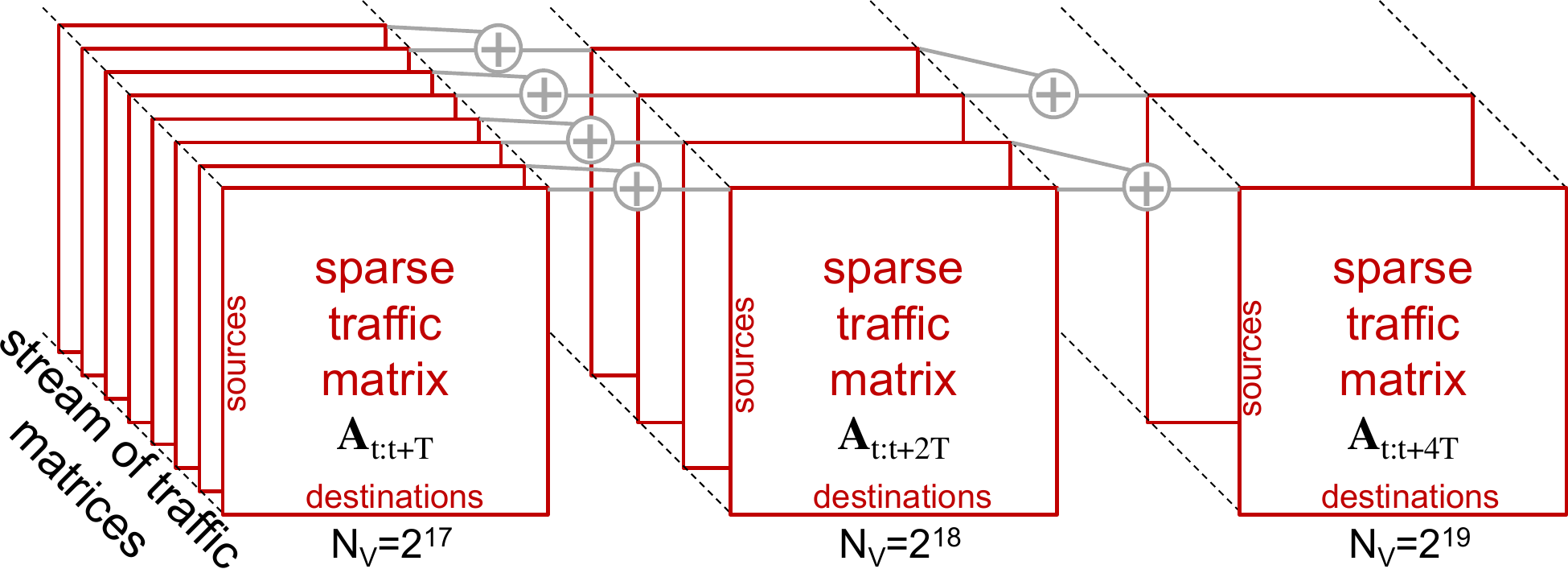}}
      	\caption{{\bf Binary Summation of Traffic Matrices.} Summing traffic matrices as binary pairs can result in more efficient memory access and more efficient analysis of matrices at each intermediate level.  Figure adapted from \cite{kepner2020multi}.}
      	\label{fig:MultiTemporalMatrix}
\end{figure}

Efficient computation of network quantities on multiple time scales can be achieved by hierarchically aggregating data in different time windows \cite{kepner2020multi}.  Figure~\ref{fig:MultiTemporalMatrix} illustrates a binary aggregation of  different streaming traffic matrices.   Computing each quantity at each hierarchy level eliminates redundant computations that would be performed if each packet window was computed separately.  Hierarchy also ensures that most computations are performed on smaller matrices residing in faster memory.  Correlations among the matrices mean  that adding two matrices each with $N_V$ entries results in a matrix with fewer than $2N_V$ entries, reducing the relative number of operations as the matrices grow.

\section{The Graph Challenge}

\begin{figure*}
\center{\includegraphics[width=2.0\columnwidth]{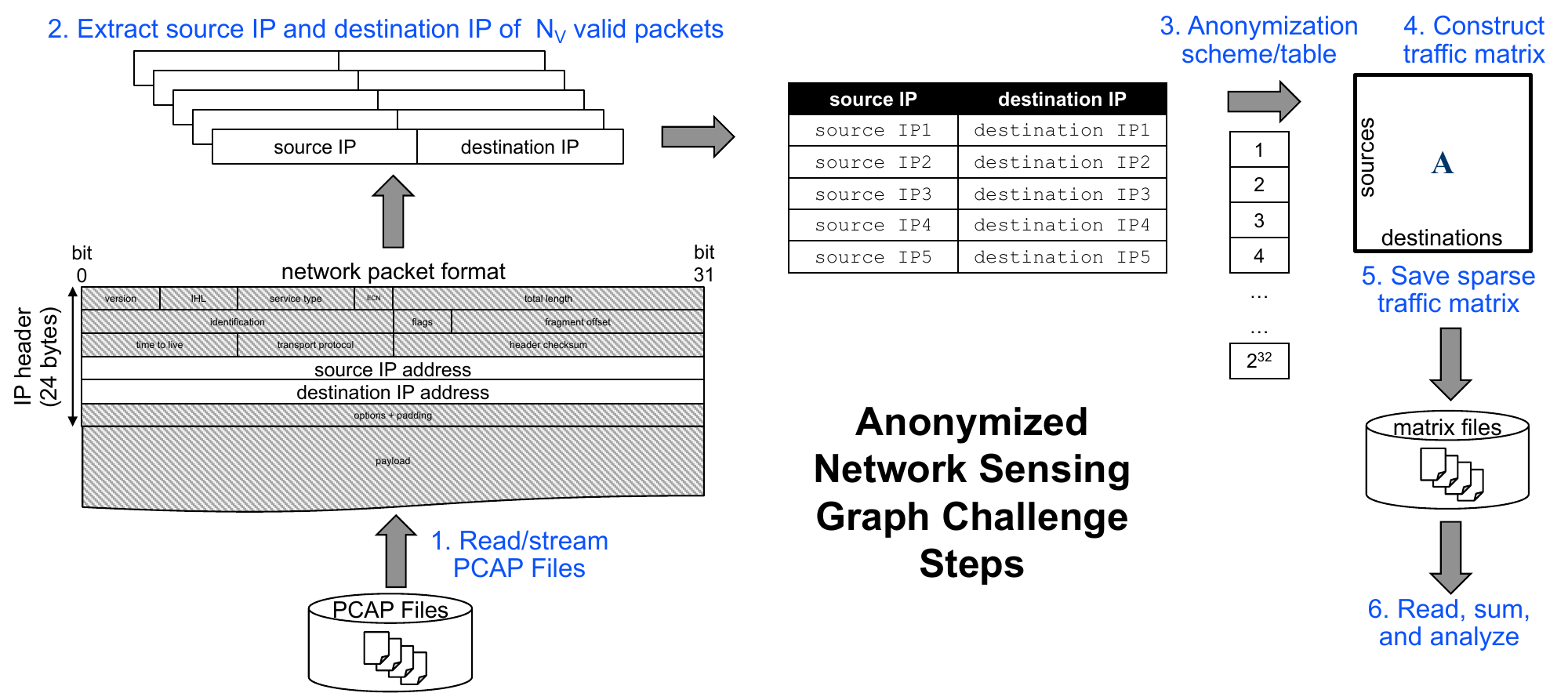}}
      	\caption{{\bf Anonymized Network Traffic Graph Challenge Steps}.
(1) Read/stream each network packet capture (PCAP) file.
(2) Extract the source IP and destination IP addresses from the packet headers and buffer $N_V$ valid packets.
(3) Anonymize the source IP and destination IP. Anonymization should  be consistent over all files so all the data can be meaningfully further aggregated. Assume that any pair in the $2^{32}{\times}{2^{32}}$ IPv4 traffic space is possible.
(4) Construct sequential traffic matrices from $N_V$ valid packets. Matrices should align with the mathematical definition of a matrix.
(5) Save the traffic matrices to files (granularity is up to the implementor).
(6) Read in the traffic matrix files, sum the traffic matrices associated with a PCAP file into one large traffic matrix ${\bf A}_t$ (see Figure~\ref{fig:MultiTemporalMatrix}), and perform the analysis highlighted in Table~\ref{tab:Aggregates}.
}
      	\label{fig:AnonNetSenseGraphChallenge}
\end{figure*}

\begin{figure}
\noindent \rule{\columnwidth}{0.5pt}

{\small\sf
\noindent AnonNetSensingGraphChallenge(

~  PCAPfile, \# name of PCAP file
  
~  Np, \# packets in file ($2^{30}$)
  
~  Nv, \# packets per matrix ($2^{17}$)
  
~  NmatPerFile, \# matrices per output file ($2^{6}$)
  
~  anonKey  \# anonymization key
  
\noindent  );

  PCAPbuffer =  read(PCAPfile);
  
  p = 0;
  
  {\bf for} i = 0 {\bf to} (Np/(NmatPerFile*Nv))-1 \# ($2^{7}-1$)

     ~ {\bf for} j = 0 {\bf to} NmatPerFile-1

     ~  ~ {\bf for} k = 0 {\bf to} Nv-1

     ~  ~  ~ [srcIP(k) destIP(k)] = readPCAPheader(PCAPbuffer,p);

     ~  ~  ~ srcIPanon(k) = anonymize(srcIP,anonKey);
     
     ~  ~  ~ destIPanon(k) = anonymize(destIP,anonKey);     
     
     ~  ~  ~ p++;
     
     ~  ~ {\bf end}

     ~  ~ ${\bf A}$[j] = constructMatrix(srcIPanon,destIPanon);

     ~ {\bf end}
     
     saveMatrices(${\bf A}$,i);
 
  {\bf end}
  
  ${\bf A}_t$(:,:) = 0;

  {\bf for} i = 0 {\bf to} (Np/(NmatPerFile*Nv))-1
    
     ~ ${\bf A}$ = readMatrices(i);
    
     ~ {\bf for} j = 0 {\bf to} NmatPerFile-1

     ~  ~ ${\bf A}_t$ += ${\bf A}$[j];
     
     ~ {\bf end}
 
  {\bf end}

  \# perform the analysis on ${\bf A}_t$ listed in Table~\ref{tab:Aggregates}

\noindent {\bf end}}

\noindent \rule{\columnwidth}{0.5pt}

      	\caption{{\bf Anonymized Network Traffic Graph Challenge Pseudocode}. Code begins by reading a $2^{30}$ packet PCAP file in groups of $N_V = 2^{17}$ valid packets.  The source and destination IP addresses of the packets are anonymized and then used to populate the entries of a traffic matrix ${\bf A}$[j]. $2^6$ of these traffic matrices are then saved as individual files within a .tar file.  $2^7$ .tar files are saved per PCAP files.  After all the traffic matrices are constructed and saved, the .tar files are then read again and all the traffic matrices are summed into one traffic matrix ${\bf A}_t$.  The analysis highlighted in Table~\ref{tab:Aggregates} are then performed on ${\bf A}_t$ and reported.}
      	\label{fig:AnonNetSenseCode}
\end{figure}

This challenge provides an opportunity to highlight novel approaches for optimizing the construction and analysis of anonymized traffic matrices.  This paper describes the anonymization, construction, and analysis of these traffic matrices.  The overall steps of the challenge are depicted in Figure~\ref{fig:AnonNetSenseGraphChallenge}.  The Anonymized Network Traffic Graph Challenge consist of several timed steps, each of which can be important to optimize in a real system.  The complete process for performing the challenge consists of the following steps

\begin{enumerate}
\item \underline{Timed}: Read/stream each network packet capture (PCAP) file containing $2^{30}$ packets.
\item \underline{Timed}: Extract the source IP address and destination IP address from each packet header.
\item \underline{Timed}: Anonymize the source IP and destination IP. Anonymization should be consistent over all files so all the data can be meaningfully further aggregated. Assume that any pair in the $2^{32}{\times}2^{32}$ IPv4 traffic space is possible. Novel approaches that also handle 128-bit IPv6 are encouraged. Anonymization can be done at different levels as long as it is explicitly stated: [trivial] no anonymization, [reference implementation] trusted sharing employing anonymization (e.g., CryptoPAN) that assumes the existence of an agreement prohibiting deanonymization, [advanced research] semantically secure anonymization.
\item \underline{Timed}: Construct sequential traffic matrices with $N_V = 2^{17}$ packets (this size is large enough for meaningful statistics but small enough to preserve enough temporal information for statistical analysis given that Internet packets can arrive in any order).  Matrices should be aligned with the mathematical definition of a matrix and can be read directly into an available matrix analysis environment.   Avoid internal redundancy and store each (i,j) pair only once. Valid matrix formats include, but are not limited to, compressed sparse rows (CSR), compressed sparse columns (CSC), and sorted triples.  Proprietary binary formats are allowed.
\item \underline{Timed}: Save the traffic matrices to files.  Valid file formants include, but are not limited to, comma separated values (CSV), tab separated values (TSV), SuiteSparse GraphBLAS [reference implementation], HDF, CDF, and NetCDF.  The number of files and number of traffic matrices per file is up to the implementor and range from $2^{13}$ files each containing 1 traffic matrix to  1 file with $2^{13}$ traffic matrices.  The reference implementation saves $2^7$ .tar files each containing $2^6$ SuiteSparse GraphBLAS traffic matrices.
\item \underline{Timed}: Read in the $2^{13}$ traffic matrices associated with $2^{30}$ packets, sum the traffic matrices into a single large traffic matrix ${\bf A}_t$ (see Figure~\ref{fig:MultiTemporalMatrix}), and perform the analysis highlighted in Table~\ref{tab:Aggregates}.
\end{enumerate}

Reference serial implementations in various programming languages are available at GraphChallenge.org. The pseudo-code for the reference implementation is shown in Figure~\ref{fig:AnonNetSenseCode}. For a given implementation of the Graph Challenge an implementor should keep the following guidance in mind.
\centerline{\underline{Do}}
\begin{itemize}
\item Use an implementation that could work on real-world data.
\item Distribute inputs and run in data parallel mode to achieve higher performance (this may require storing traffic matrices on every processor and increase the memory footprint).
\item Split up steps and run in a pipeline parallel mode to achieve higher performance (this saves memory, but requires communicating results after each group of steps).
\item Use other reasonable optimizations that would work on real-world data.
\end{itemize}
\centerline{\underline{Avoid}}
\begin{itemize}
\item Using optimizations that would not work on real-world data.
\item Unnecessarily pre-computing quantities for a subsequent step in a previous step.
\end{itemize}

\section{Computational Metrics}
\label{sec:metrics}
Submissions to the Anonymized Network Sensing Graph Challenge will be evaluated on the overall innovations highlighted by the implementation and two metrics: correctness and performance.

\subsection{Correctness}
\label{sec:correctness}
Correctness is evaluated by comparing the reported Table~\ref{tab:Aggregates} quantities for each $2^{30}$ packet PCAP file with the ground truth provided.

\subsection{Performance}
\label{sec:perf}
The performance of the algorithm implementation should be reported in terms of the following metrics.
\begin{itemize}
\item Total number of packets processed: this measures the amount of data processed.
\item Execution time: total time required to perform the Graph Challenge.
\item Rate: measures the throughput of the implementation as the ratio of the number of packets processed divided by the execution time.
\item Processor: number and type of processors used in the computation.
\end{itemize}
Graph Challenge participants are free to select (with accompanying explanation) the elements of any of the Graph Challenges that are appropriate for highlighting their innovations.  Reporting the performance of individual steps in the Graph Challenge are encouraged.  It is often the case that a particular innovation is focused on improving a single step.

\section{Anonymized Test Data}

\begin{figure}
\center{\includegraphics[width=1.0\columnwidth]{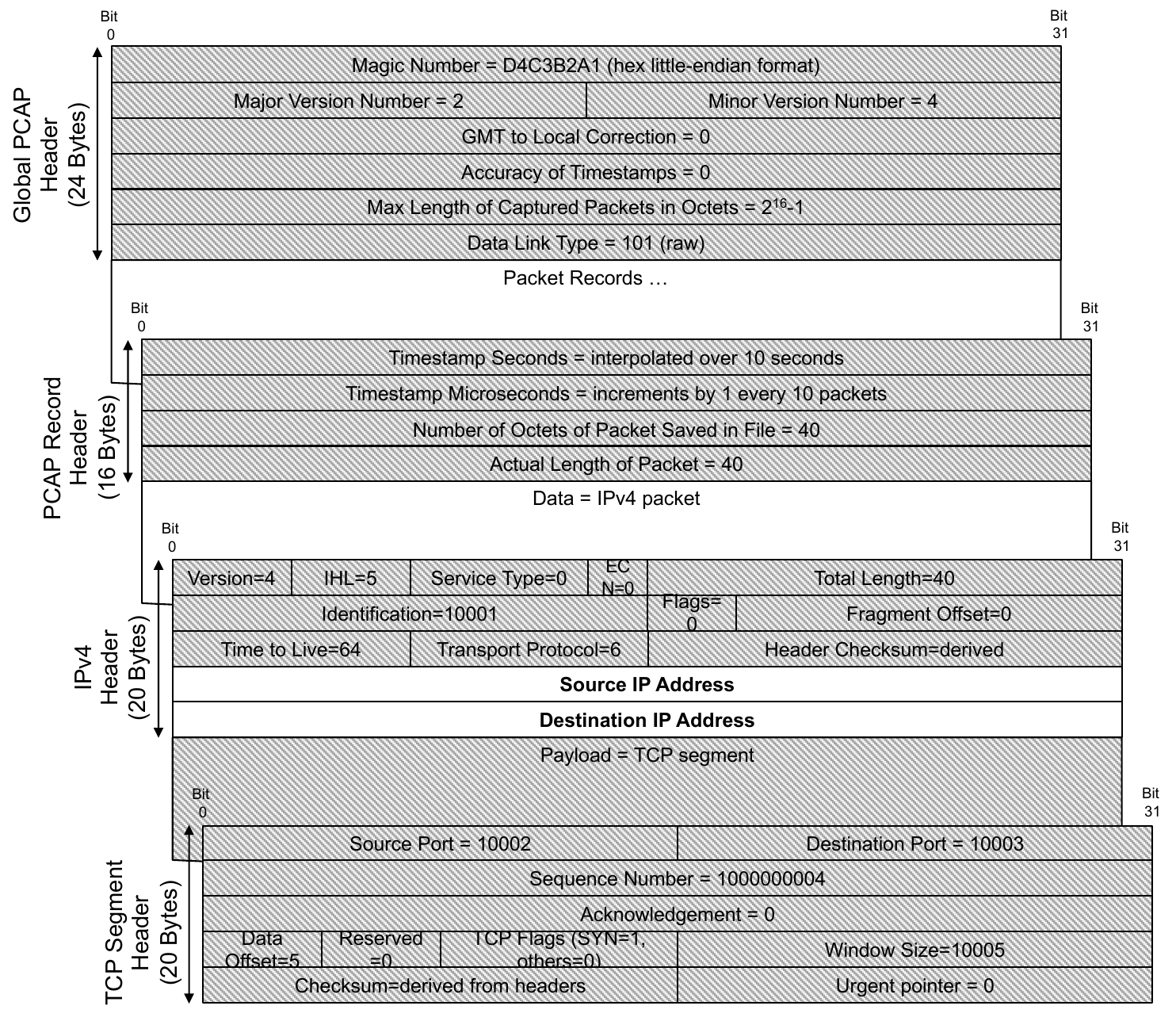}}
      	\caption{{\bf Packet Capture (PCAP) File Format.}  Each of $2^{30}$ packets are stored in a PCAP file in the above format.  PCAP files begin with a global header followed by a sequence of PCAP records.  Each PCAP record consists of a header and data containing an  Internet Protocol version 4 (IPv4) packet.  An IPv4 packet consists of a header containing the source address and the destination address fields.  For completeness, the IPv4 payload contains a Transmission Control Protocol (TCP) segment header.  The specific values or method used to populate the fields are listed in the figure. [Note: setting data link type to 101 indicates this a raw data and there is no Ethernet frame header between the PCAP record header and the IPv4 header.]}
      	\label{fig:PCAPfileformat}
\end{figure}

The test data consists of two types (1) randomized and (2) anonymized real data derived the from the CAIDA telescope.  The randomized data consists of a single freely available $2^{30}$ packet PCAP file with source and destination IP addresses generated with $2{\times}2^{30}$ calls of the C PCG32 (Permuted Congruential Generator 32 bit) pseudo random number generator function \cite{oneill:pcg2014}.  The CAIDA darknet telescope is a significant portion of a globally routed /8 network carrying essentially no legitimate traffic since it is an Internet darkspace, providing an ideal vantage point by which to observe and study unsolicited anomalous traffic.  The CAIDA network traffic is collected and anonymized into traffic matrices in a process similar to steps 1-5 shown in Figure~\ref{fig:AnonNetSenseGraphChallenge}\cite{jones2023deployment}.  Subsets of these traffic matrices representing $2^{30}$ contiguous packets were selected around noon and midnight from many days around the first quarter of 2022 (see Table~\ref{tab:DataSets}) \cite{jananthan2023mapping}.  These traffic matrices were then converted back into $2^{30}$ packet PCAP files.  For both the random and CAIDA data, the other fields in the PCAP header are populated using the values or methods shown in Figure~\ref{fig:PCAPfileformat}.

An additional enrichment data set is also included that looks up sources found in the CAIDA telescope data in the GreyNoise honeyfarm database \cite{kawaminami2022large,jananthan2023mapping,Greynoise2024}. The GreyNoise honeyfarm is made up of thousands of servers carrying out conversations with sources scanning the Internet; based on these conversations GreyNoise can associate various metadata with those sources to collectively build a refined picture of the malicious sources regularly scanning the Internet and the techniques they employ.   The GreyNoise enrichment of CAIDA data uses anonymized IP addresses throughout and is provided as a point of departure for future investigations (see \cite{kawaminami2022large,jananthan2023mapping} for details on the enriched fields provided).

\begin{table}
\caption{Anonymized Data Sets}
\vspace{-0.25cm}
Characteristics of CAIDA Telescope derived anonymized PCAP files and corresponding GreyNoise enrichment data.
\begin{center}
\begin{tabular}{cccccc}
\hline
                       & \textbf{CAIDA} & \textbf{CAIDA data size} & \textbf{GreyNoise} \\
\textbf{Month} & \textbf{$2^{30}$ packet sets} & \textbf{compressed} & \textbf{data size} \\
\hline
2022-01 &  25 (noon); 24 (midnight) & 375 GB  & 3.6 GB \\
\hline
2022-02 &  17 (noon); 18 (midnight) & 271 GB  & 3.6 GB  \\
\hline
2022-03 &  24 (noon); 26 (midnight) & 432 GB & 3.6 GB \\
\hline
2022-04 &  13 (noon); 14 (midnight) &  242 GB &   \\
\hline
\end{tabular}
\end{center}
\label{tab:DataSets}
\end{table}

\section{Performance Measurements}
\begin{figure}
\center{\includegraphics[width=1.0\columnwidth]{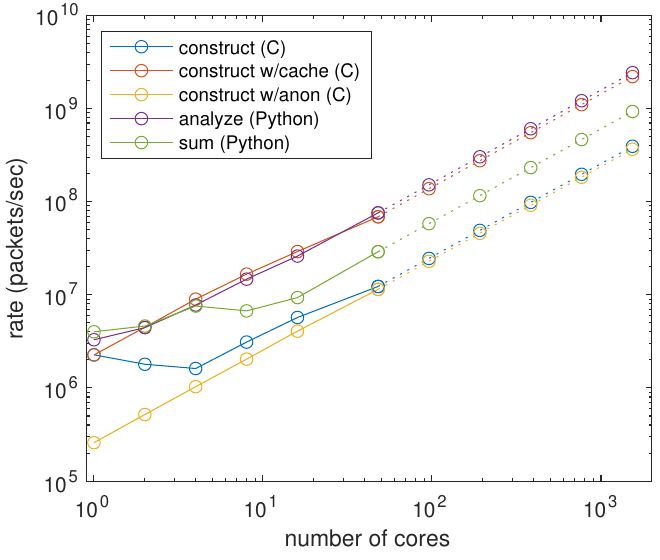}}
      	\caption{{\bf Reference Performance.} Average performance measurements of the reference C traffic matrix constructor code (with anonymization and with file caching) and the Python traffic matrix sum and analyze code.  Effective bandwidth is computed assuming 10,000 bits/packet for real packet data on a real network.  The constructor code has a small memory footprint and 48 distinct instances each processing a separate PCAP file can be run on a 48 core node.  The sum and analyze code have a larger memory footprint and 3 distinct instances each with 16 OpenMP threads each processing a separate PCAP file can be run on a 48 core node with 192 GB of RAM.  The multi-node performance scales linearly over distinct PCAP files.}
      	\label{fig:ReferencePerformance}
\end{figure}

  Performance measurements of the reference C traffic matrix constructor code and the Python traffic matrix sum and analysis code are shown in Figure~\ref{fig:ReferencePerformance} and provide one example for reporting results.  Parallel implementations of the reference code were developed and tested using dual 2.4 GHz Intel Xeon Platinum 8260 processor compute nodes on the MIT SuperCloud TX-Green  supercomputer \cite{reuther2018interactive}.   These results demonstrate that traffic matrix construction can be done in a streaming fashion with modest memory enabling large numbers of PCAP files to be processed simultaneously on a single compute node.  Reading the PCAP files can take significant time as shown by the rate increase achieved by caching the files in memory.  Likewise, in-line anonymization has the opposite effect.  Prior work shows that anonymization time can be effectively eliminated by using look-up tables \cite{jones2022graphblas}. Sum and analysis of the traffic matrices requires a larger memory footprint which can be accelerated with threads.  In both cases, multiple files can be processed simultaneously and the performance scales linearly with nodes.

\section{Conclusions}

The anonymized network sensing Graph Challenge seeks to enable large, open, community-based approaches to protecting networks.    Community access to very broad data sets with the highest regard for privacy is essential for solving many large-scale networking problems. Anonymized source-to-destination traffic matrices with standard data sharing agreements have emerged in the broader networking community as a data product that can meet many of these requirements.  Using over 100 billion network packets derived from the largest Internet telescope in the world (CAIDA) the anonymized network sensing Graph Challenge provides an opportunity to highlight novel approaches for optimizing the construction and analysis of anonymized traffic matrices.  A GraphBLAS reference implementation is provided, but the use of GraphBLAS is not required in this Graph Challenge. As with prior Graph Challenges the goal is to provide a well-defined context for demonstrating innovation. Graph Challenge participants are free to select (with accompanying explanation) the elements of any of the Graph Challenges that are appropriate for highlighting their innovations.

\section*{Acknowledgments}

The authors wish to acknowledge the following individuals for their contributions and support: Daniel Andersen, LaToya Anderson, Sean Atkins, David Bader, Chris Birardi, Bob Bond, Alex Bonn, Koley Borchard, Stephen Buckley, Aydin Buluc, K Claffy, Cary Conrad, Chris Demchak, Phil Dykstra, Alan Edelman, Peter Fisher, Garry Floyd, Jeff Gottschalk, Dhruv Gupta, Oded Green, Thomas Hardjono, Chris Hill, Miriam Leeser, Charles Leiserson, Chris Long, Kirsten Malvey, Sanjeev Mohindra, Roger Pearce, Heidi Perry, Ali Pinar, Christian Prothmann, Steve Rejto, Josh Rountree, Daniela Rus, Siddharth Samsi, Mark Sherman, Scott Weed, Michael Wright, Marc Zissman.

\bibliographystyle{ieeetr}
\bibliography{AnonNetSensing}

\end{document}